\documentclass[12pt]{iopart}
\usepackage{iopams} 
\usepackage[utf8]{inputenc}
\usepackage{graphicx}
\usepackage{slashed}
\begin{document}

\begin{flushright}
HIP-2017-18/TH \\
\end{flushright}
\vspace*{5mm}

\title{Collisionless shocks in self-interacting dark matter}

\author{Matti Heikinheimo}
\address{Helsinki Institute of Physics and Department of Physics, University of Helsinki, Finland}
\ead{matti.heikinheimo@helsinki.fi}
\author{Martti Raidal, Christian Spethmann, Hardi Veerm\"ae}
\address{National Institute of Chemical Physics and Biophysics, Tallinn, Estonia}

\begin{abstract}
Self-interacting dark matter has been proposed as a solution to small scale problems in cosmological structure formation, and hints of dark matter self scattering have been observed in mergers of galaxy clusters. One of the simplest models for self-interacting dark matter is a particle that is charged under \emph{dark electromagnetism}, a new gauge interaction analogous to the usual electromagnetic force, but operating on the dark matter particle instead of the visible particles. In this case, the collisional behaviour of dark matter is primarily due to the formation of collisionless shocks, that should affect the distribution of DM in merging galaxy clusters. We evaluate the time and length scales of shock formation in cluster mergers, and discuss the implications for modelling charged dark matter in cosmological simulations.
\end{abstract}
\noindent{\it Keywords\/}: Self-interacting dark matter, cluster mergers, collisionless shocks.

\section{Introduction}
Galaxies, such as the Milky Way, are embedded in haloes of invisible particles that do not emit light nor scatter from the ordinary nuclear matter. This unknown substance, called Dark matter (DM), is five times as abundant in the universe as the ordinary visible matter~\cite{Ade:2015xua}, as is observed from galactic rotation curves, gravitational lensing observations of galaxy clusters, and from the baryon acoustic oscillations present in the cosmic microwave background and in the large scale structure of the universe. However, the Standard Model (SM) of particle physics does not contain a suitable dark matter candidate, and thus an explanation is required from physics beyond the SM.

The simplest, most studied models of DM are extensions of the SM that contain an additional weakly interacting massive particle (WIMP), that was once in thermal equilibrium with the SM in the early universe. The abundance of DM in these models is produced as a thermal relic, as the co-moving number density freezes out when the expansion rate of the universe overtakes the annihilation rate of the DM particles into SM species. For details, we refer the reader to the large body of literature on the WIMP paradigm, as reviewed e.g. in~\cite{Bergstrom:2000pn,Bertone:2016nfn,deSwart:2017heh,Arcadi:2017kky}.

The WIMP DM is characterized by the electroweak scale mass and interaction strength, implying that today it exists in a form of cold, collisionless and pressurless dust, only experiencing the effects of gravity. This model can very accurately describe the formation of cosmic structure at large scales, but at subgalactic scales some discrepancies appear between the numerical simulations of structure formation and the observed small scale structure of our cosmic environment. These issues are related to the steepness of the DM halo density profile, the so called cusp versus core problem~\cite{Moore:1994yx}, and to the abundance of DM substructures at scales below the size of a typical galaxy, the so called missing satellites problem~\cite{Klypin:1999uc}. While it is possible that the small scale structure problems may be alleviated by carefully accounting for the effects of baryons in the numerical simulations~\cite{Fattahi:2016nld}, it has been proposed that self-interacting DM can resolve both issues~\cite{Spergel:1999mh,Tulin:2017ara}. Self-interacting DM is a particle species that interacts weakly with the SM particles, but exhibits substantial self-scattering of the order of \mbox{$\sigma/m_{\rm DM}\sim 1\ {\rm cm}^2/{\rm g}$}, where $\sigma$ is the elastic self-scattering cross section and $m_{\rm DM}$ is the mass of the dark matter particle. This interaction strength is similar to the low-energy scattering between protons and neutrons in the SM, and is 14 orders of magnitude larger than the annihilation cross section required for a thermal relic WIMP with a typical mass of the order of 100 GeV.

An intriguing way of observing the dynamical properties of DM is provided by dissociative mergers, where two galaxy clusters have passed through each other in the initial stage of the merger. These events are called dissociative, because the intracluster medium (ICM), consisting of hot hydrogen plasma and visible in X-rays, is shocked and slowed down during core passage, while the luminous stars in galaxies pass through without collisions, and thus the X-ray emitting ICM and optically visible stars become displaced from each other. The spatial distribution of DM in these structures can be determined via gravitational lensing. It is generally assumed that if DM behaves as a collisionless dust, as within the WIMP paradigm, it will remain coincident with the visible stars, whereas self-interacting DM is believed to exhibit collisional behaviour and lag behind the stars, creating an offset between the center of gravitational mass and the optical luminosity peak~\cite{Kahlhoefer:2013dca,Kim:2016ujt,Ng:2017bwr}. Dissociative mergers are therefore considered a promissing avenue for directly probing the self-interactions of DM.

Currently only a few dissociative mergers have been observed~\cite{Harvey:2015hha} with enough accuracy to reconstruct a gravitational lensing map of the DM, and the status of self-interacting DM remains unclear. While no compelling evidence for the offsets between DM and luminous stars has been observed, the constraining power of these observations have been guestioned in the recent literature~\cite{Wittman:2017gxn,Robertson:2016xjh}. Additionally, tentative evidence for an offset between the stars and the DM halo of an infalling galaxy in the center of the cluster Abell 3827 has been observed~\cite{Massey:2015dkw,Kahlhoefer:2015vua,Taylor:2017ipx} and interpreted as evidence for DM self-interactions. In the dissociative merger Abell 520, a DM substructure coincident with the shocked ICM has been observed~\cite{Mahdavi:2007yp,Jee:2012sr,Jee:2014hja}. We will argue that this puzzling situation may be explained by a multi-component model of DM, with a dominant collisionless WIMP-like component, and a subdominant self-interacting component, that exhibits similar collisional behaviour as the ICM, and is therefore named dark plasma~\cite{Heikinheimo:2015kra}.

The structure of this paper is as follows: In section \ref{cluster mergers} we will describe two well known dissociative mergers, the Bullet cluster (1E 0657-558) and the "train wreck" cluster \mbox{Abell 520}, and motivate our two-component DM matter model with dark plasma. In section \ref{dark plasma} we will discuss the idea of a charged DM species, and discuss the role of plasma physics in understanding the behaviour of such substance. In section \ref{conclusions} we conclude our discussion with a future outlook.

\section{Dissociative cluster mergers}
\label{cluster mergers}
Perhaps the most well known dissociative merger is the Bullet cluster, 1E 0657-558. In this system a smaller subcluster (the bullet) of $\sim 2\times 10^{14}\ {\rm M}_\odot$ has passed through a ten times more massive cluster with the peak relative velocity of $\sim 4000\ {\rm km}/{\rm s}$~\cite{Springel:2007tu,Lage:2013yxa}. The ICM forms a bow shock in the center of the system, while the luminous galaxies are further away, with the luminosity peaks of the two substructures separated by $\sim 720$ kpc. The weak lensing reconstruction of the mass distribution shows the DM coincident with the luminous galaxies within the observational margin of error, and clearly separated from the shocked ICM~\cite{Clowe:2006eq}. The non-observation of an offset between the DM center of mass and the luminosity peak of the galaxies has been interpreted as suggestive of the non-collisional nature of DM, and has been used to constrain the self-interaction cross section at the level of \mbox{$\sigma/m_{\rm DM}\lesssim 1\ {\rm cm}^2/{\rm g}$}~\cite{Randall:2007ph}, although this limit has been called to question in more advanced simulations~\cite{Robertson:2016xjh}. Another method of constraining the self-interaction strength is found by observing the mass-to-light ratio of the bullet substructure, leading to a conclusion that the subcluster can not have lost more than 30\% of its mass as it passed through the main halo~\cite{Markevitch:2003at}. This observation has been argued to place a similar constraint of the self-interaction cross section, of the order of $1\ {\rm cm}^2/{\rm g}$, in the case that all of dark matter is of the self-interacting type.

The Abell 520 cluster exhibits a more complicated structure, and is belived to be a merger of multiple smaller substructures, with the main merger driven by two $\sim 4\times 10^{13}\ {\rm M}_\odot$ subclusters~\cite{Mahdavi:2007yp} that have passed through each other with the relative velocity of 2300 km/s~\cite{Markevitch:2004qk}, as indicated by the bow shock visible in the X-ray images of the ICM. The total mass of this system is estimated as $\sim 10^{15}\ {\rm M}_\odot$. What makes this system peculiar, is the presence of an exess of non-luminous matter in the weak lensing images of the cluster, coincident with the shocked ICM~\cite{Mahdavi:2007yp,Jee:2012sr,Jee:2014hja}. The mass of this stucture is similar to the masses of the two main subclusters. Explaining the presence of this substructure with self-interacting DM was estimated to require a self-interaction cross section of the order of 3.8 cm$^2$/g~\cite{Mahdavi:2007yp}, clearly at odds with the constraints from other sources.

In order to reconcile the seemingly contradicting observations of the Bullet cluster and Abell 520, we have proposed a model of two-component DM~\cite{Heikinheimo:2015kra}, where the main component is composed of a collisionless WIMP, and a subdominant fraction that contributes some 25-30\% of the abundance of DM is made up of a self-interacting species that behaves like an effectively collisional fluid in the cluster mergers. In the Abell 520 merger of roughly equal size clusters, the self-interacting component is shocked and separated from the luminous galaxies and the collisionless DM component, forming the observed dark core coincident with the shocked ICM. In the Bullet cluster the self-interacting component of the smaller subcluster is mostly absorbed by the halo of the larger subcluster during core passage, and no additional dark core is seen. This qualitative picture has been reproduced in a hydrodynamical simulation~\cite{Sepp:2016tfs} of the two dissociative mergers with a 25\% fraction of dark matter in the form of a collisional fluid-like component.

\section{Dark plasma and collisionless shocks}
\label{dark plasma} 
As discussed above, the self-interacting DM component in our model should effectively behave as a collisional fluid in the scales relevant for cluster mergers. If this behaviour should be caused by hard binary collisions of the DM particles, this would require a self-interaction cross section in the range \mbox{$\sigma/m_{\rm DM}\gtrsim 10\ {\rm cm}^2/{\rm g}$}~\cite{Spergel:1999mh}, which is enormous compared to the annihilation cross section required from a thermal relic, unless the DM particle is very light. Instead, we have proposed a simple model with a DM particle that is charged under a new $U(1)$ interaction, reminiscent to the electromagnetic force in the SM, but operating in the hidden sector. This particle physics model is described by the Lagrangian
\begin{equation}
\mathcal{L}_{\rm D} = \frac{1}{4} F_{{\rm D}\mu\nu}F^{\mu\nu}_{\rm D} + \bar{\chi}\left(i \slashed{D}  - m_{\rm D}\right)\chi,
\label{Lagrangian}
\end{equation}
where $D^\mu = \partial^\mu - e_{\rm D} A_{\rm D}^\mu$ is the covariant derivative, $F_{\rm D}^{\mu\nu}$ is the field strength tensor of the dark photon $A^{\mu}_{\rm D}$, $\chi$ is the interacting DM particle with mass $m_{\rm D}$ and $e_{\rm D}$ is the dark $U(1)$ charge. We work in the limit where the kinetic mixing term $F^{\rm D}_{\mu \nu} F^{\mu \nu}$ is set to zero, since such a term is severely constrained by recombination and halo dynamics \cite{McDermott:2010pa}. As discussed in~\cite{Heikinheimo:2015kra}, a $\sim 30\%$ fraction of DM in the form of this self-interacting species can be produced as a thermal relic if the dark charge and mass of the DM particle $\chi$ are related by
\begin{equation}
\alpha_{\rm D}\approx 10^{-4}\frac{m_{\rm D}}{\rm GeV},
\label{alpha vs mD}
\end{equation}
where $\alpha_{\rm D} = e_{D}^{2}/4\pi$ is the fine structure constant of the dark $U(1)$, and we work in natural units where $\hbar=c=1$.  Models like this one, where DM or a component of it is charged under a new long-range electromagnetism-like force, are referred to as charged dark matter, and various aspects of these models have been studied in the literature~\cite{Foot:2004pa,Feng:2008mu,Ackerman:mha,Feng:2009mn,Kaplan:2009de,Fan:2013yva,Agrawal:2016quu,Rosenberg:2017qia}. However, the effects of collisionless shocks, which according to our view are responsible for the effective collisional fluid-like behaviour of the charged DM component at cluster scales, have not received much attention in the above references.

Collisionless shocks and their formation mechanisms have been described in~\cite{Bret:2013qva,Bret:2015qia}. They arise as the opposing bodies of plasma begin to overlap and counter-stream. In the counter-streaming region electromagnetic instabilities begin to grow, until the magnetic fields created by the instabilities become strong enough to deflect the incoming stream of charged particles. The overlapping region becomes impenetrable and a discontinuity in the velocity field of the plasma forms, creating a shock wave. These shocks will dissipate energy and thermalize the plasma streams, so that at scales much larger than $L_{\rm s}=2v_{\rm rel}\tau_{\rm s}$, where $v_{\rm rel}$ is the relative velocity of the colliding bodies of plasma and $\tau_{\rm s}$ is the shock formation time, the system behaves effectively like a collisional fluid~\cite{Bret:2015qia}.

The shock formation time is inversely proportional to the plasma frequency~\cite{Bret:2013qva}, $\tau_{\rm s}\sim \omega_{\rm p}^{-1}$, where
$\omega_{\rm p}^2 = 4\pi\alpha_{\rm D} n_{\rm D}/m_{\rm D}$ and $n_{\rm D}$ is the number density of the charged DM particles. The average mass density of DM within a subcluster in a dissociative merger such as the Abell 520 is approximately $\sim 10^{-2}\ {\rm GeV}/{\rm cm}^3$~\cite{Heikinheimo:2015kra}, so that substituting equation (\ref{alpha vs mD}) into the above yields an estimate for the inverse plasma frequency:
\begin{equation}
1/\omega_{\rm p} \approx 60\ {\rm ms}\ \sqrt{\frac{m_{\rm D}}{\rm GeV}}.
\end{equation}
A conservative order of magnitude estimate for the shock formation time scale is thus given by
\begin{equation}
\tau_{\rm s} \sim 10^3\omega_{\rm p}^{-1}\approx 60\ {\rm s}\ \sqrt{\frac{m_{\rm D}}{\rm GeV}}.
\end{equation}
The corresponding lenght scale for a typical relative velocity in a dissociative merger $v_{\rm rel}\sim 10^3\ {\rm km}/{\rm s}$ is then $L_{\rm s}= 2v_{\rm rel}\tau_{\rm s}\sim 10^5\ {\rm km}\sim 10^{-9}\ {\rm pc}$. This length scale is many orders of magnitude below the resolution of any cosmological N-body/hydrodynamical simulation, so that the fluid approximation for the dark plasma in these simulations should be valid. The conclusion is that a charged DM component, such as the one described by the particle physics model of equation (\ref{Lagrangian}), will behave like a collisional fluid at scales relevant for cluster mergers. For comparison, the mean free path of the charged DM particles due to binary collisions in a merger such as Abell 520 was approximated~\cite{Heikinheimo:2015kra} as $\lambda_f\approx 40\ {\rm kpc}\ (m_{\rm D}/{\rm GeV})$, confirming that in the sense of binary collisions the dark plasma indeed is collisionless in the scales where the shock formation takes place, and the effective fluid-like behaviour at larger scales is governed by the formation of collisionless shocks.

\section{Conclusions and outlook}
\label{conclusions}
Self-interacting DM has been proposed as a solution to the problems that seem to appear at small scale structure formation if DM is assumed to be collisionless. Self-interactions of DM particles may be directly probed by observing mergers of cosmic structures, such as galaxies and clusters. While some observations, most notably the Bullet cluster, seem compatible with the collisionless DM picture, the minor merger in Abell 3827 and the major dissociative merger in Abell 520 clusters seem to exhibit behaviour that could be associated with self-scattering of DM particles.

Charged DM is a simple and well motivated model for self-interacting dark matter, and different variations of this model have been widely studied. Observational constraints for the parameters of such models have been derived~\cite{Agrawal:2016quu} from considerations of the effects of DM self-interactions on the ellipticity of DM haloes, the stripping of self-interacting DM from small satellite galaxy haloes as they pass through the larger halo of the host galaxy, and from observations of the dissociative mergers discussed above. However, the effects of collisionless shocks have been mostly overlooked in these considerations. As we have demostrated with simple estimates of the shock formation time and length scales, the effective collisional behaviour of charged DM should be dominantly dictated by these processes, potentially leading to much stronger constraints on the parameters of the model than what is depicted in the current literature.

Understanding the cosmological and astrophysical consequences of self-interacting DM relies largely on N-body hydrodynamical simulations, where the self-interacting DM may be treated as a fluid~\cite{Sepp:2016tfs}, in case that the microphysical processes are effective in bringing the system into local thermal equilibrium at scales below the resolution of the simulation. In the opposing optically thin regime, where the self-interactions are not effective enough to bring the system into local thermal equilibrium at sub-grid scales, the self-scattering may be implemented as a stochastic process as described e.g. in~\cite{Robertson:2016qef}. When the self-interactions are dominantly point-like hard scattering processes the identification of these two regimes is quite straightforward, but the situation is less clear in the case of charged DM, where the microphysics is dominated by plasma instabilities. To facilitate a more comprehensive study of the collisional phenomenology of charged DM, an important piece of information would be to understand in more detail the exact scale above which the pair plasma may be effectively treated as a collisional fluid, and how this scale depends on the parameters of the model, i.e. the mass, charge and number density of the charged DM particles.

\section*{Acknowledgments}
The work of MH has been supported by the Academy of Finland, grant \# 267842.

\end{document}